\documentclass[proof]{pasj01}
\usepackage{lineno}
\draft 

\Received{$\langle$reception date$\rangle$}
\Accepted{$\langle$acception date$\rangle$}
\Published{$\langle$publication date$\rangle$}
\SetRunningHead{Sub-mm Water Maser in NGC 1052}{HIDOI}
\begin{document}

\title{ALMA Detection of 321 GHz water maser emission in the radio galaxy NGC 1052}
\author{Seiji \textsc{Kameno},\altaffilmark{1,2} %
        Yuichi \textsc{Harikane},\altaffilmark{3} %
        Satoko \textsc{Sawada-Satoh},\altaffilmark{4}
        Tsuyoshi \textsc{Sawada},\altaffilmark{1,2} %
        Toshiki \textsc{Saito},\altaffilmark{2} %
        Kouichiro \textsc{Nakanishi},\altaffilmark{2} and %
        Elizabeth \textsc{Humphreys},\altaffilmark{1} %
}

\altaffiltext{1}{Joint ALMA Observatory, Alonso de C\'{o}rdova 3107 Vitacura, Santiago 763-0355, Chile}
\altaffiltext{2}{National Astronomical Observatory of Japan, 2-21-1 Osawa, Mitaka, Tokyo 181-8588, Japan}
\altaffiltext{3}{Institute for Cosmic Ray Research, The University of Tokyo, 5-1-5 Kashiwanoha, Kashiwa, Chiba 277-8582, Japan}
\altaffiltext{4}{Graduate School of Science, Osaka Metropolitan University, 1-1 Gakuen-cho, Naka-ku, Sakai, Osaka, 599-8531, Japan}

\email{seiji.kameno@alma.cl}


\maketitle

\begin{abstract}
The Atacama Large Millimeter/submillimeter Array (ALMA) serendipitously detected H$_2$O $J_{Ka, Kc} = 10_{2,9} - 9_{3,6}$ emission at 321 GHz in NGC 1052.
This is the first submillimeter maser detection in a radio galaxy and the most luminous 321-GHz H$_2$O maser known to date with the isotropic luminosity of 1090 $L_{\Sol}$.
The line profile consists of a broad velocity component with FWHM $= 208 \pm 12$ km s$^{-1}$ straddling the systemic velocity and a narrow component with FWHM $= 44 \pm 3$ km s$^{-1}$ blueshifted by 160 km s$^{-1}$.
The profile is significantly different from the known 22-GHz $6_{1,6} - 5_{2,3}$ maser which shows a broad profile redshifted by 193 km s$^{-1}$.
The submillimeter maser is spatially unresolved with a synthesized beam of $0^{\prime \prime}.68 \times 0^{\prime \prime}.56$ and coincides with the continuum core position within 12 pc.
These results indicate amplification of the continuum emission through high-temperature ($>1000$ K) and dense ($n({\rm H}_2{\rm O}) > 10^4$ cm$^{-3}$) molecular gas in front of the core. 
\end{abstract}

\section{Introduction}
Extragalactic H$_2$O masers (megamasers) are known to originate in sub-pc-scale disks rotating around supermassive black holes in active galactic nuclei (AGNs) \citep{1995Natur.373..127M,1995PNAS...9211427M,2022NatAs...6..885I}.
Thanks to their compact and bright nature, megamasers allow $\mu$as-resolution VLBI observations \citep{2022NatAs...6..976B} making them powerful probes to weigh black-hole mass, to trace accretion matter, and to measure geometrical distance and constrain cosmological parameters \citep{2013ApJ...775...13H}.
To date 150 AGNs are known to harbor the $J_{Ka, Kc} = 6_{1,6} - 5_{2,3}$ maser at 22 GHz \citep{2018IAUS..336...86B}.

The radio galaxy NGC 1052 is a unique megamaser found in an elliptical galaxy with double-sided sub-relativistic jets \citep{1994ApJ...437L..99B,1998ApJ...500L.129C}, while the majority of megamasers are radio-quiet AGNs in spiral galaxies.
NGC 1052 shows the 22-GHz maser spectrum spanning $\sim 400$ km s$^{-1}$ \citep{1998ApJ...500L.129C,2003ApJS..146..249B,2005ApJ...620..145K} which is smooth and broad as for radio-loud minorities such as TXS 2226-184 \citep{1995Natur.378..697K}, Mrk 348 \citep{2003ApJ...590..149P} and Centaurus A \citep{2013ApJ...771L..41O},  unlike archetypal megamasers which display narrow spikes $<20$ km s$^{-1}$ spread over $\sim \pm 1000$ km s$^{-1}$ tracing (sub-)Keplerian rotation \citep{1993Natur.361...45N,2008ApJ...672..800H}.
H$_2$O masers in Mrk 348 and Centaurus A are interpreted to be excited in shocked molecular gas by interaction with jets \citep{2003ApJ...590..149P,2013ApJ...771L..41O}.
The maser spots in NGC 1052 appear to have high brightness temperature exceeding $10^9$ K and lie along jets \citep{1998ApJ...500L.129C}.
\citet{2008ApJ...680..191S} revealed the spatial distribution of maser spots consisting of two clusters on the line of sight towards the continuum components in the eastern approaching side and western receding side of jets, and found positional coincidence of maser spots and the plasma torus which obscures the core by free--free absorption (FFA) casting a gap between two maser clusters \citep{2001PASJ...53..169K,2003PASA...20..134K,2003A&A...401..113V,2004A&A...426..481K}.
As the FFA opacity decreases at higher frequency, the gap disappears above 43 GHz, and the core takes over the steep-spectrum jets at higher frequency than 86 GHz \citep{2016ApJ...830L...3S,2016A&A...593A..47B,2019A&A...623A..27B,2019ApJ...872L..21S}.
In this context, submillimeter (sub-mm) observations are crucial to probe inside the obscuring torus.

The H$_2$O $10_{2,9} - 9_{3,6}$ maser at 321.225677 GHz is a new probe for AGNs.
Excitation of the 321-GHz maser, with the lower energy level of 1846 K in the vibrational ground state, requires high temperature and high density for an inverted excitation condition.
\citet{2016MNRAS.456..374G} calculated the excitation condition of possible H$_2$O maser transitions for evolved star environments and clarified that the 321-GHz maser requires a physical temperature of $T_k > 1000$ K and high density of $n({\rm H}_2{\rm O}) > 10^4$ cm$^{-3}$.
The 321-GHz maser condition implies closer location to a central engine than that of the 22-GHz maser, if found in an AGN.

\citet{2013ApJ...768L..38H} surveyed five type-2 Seyfert galaxies selected by strong 22-GHz maser emission and discovered the first 321-GHz megamaser in the Circinus galaxy, followed by the second detection in NGC 4945 \citep{2016ApJ...827...68P, 2016ApJ...827...69H}.
\citet{2019asrc.confE..24B} referred the third detection in NGC 5643 by \citet{PesceInPrep}.
So far, no radio-loud AGN has been identified as a 321-GHz maser source.

In this letter we present the first discovery of the 321-GHz maser in NGC 1052, coupled with monitoring of the 22-GHz maser to compare the spectral profiles. Since the monitoring was taken 16 years earlier than the 321-GHz observation, we do not attempt to perform a detailed comparison however.
We employ the systemic velocity of $V_{\rm sys} = 1492$ km s$^{-1}$ and the luminosity distance of $D_L = 17.6$ Mpc.

\section{Observations}
Reduction scripts and reduced data are available in the GitHub repository\footnote{https://github.com/kamenoseiji/ALMA-2021.1.00341.S}.

\subsection{GBT}
We observed the 22-GHz maser at 6 epochs; 2005-11-18, 2005-11-28, 2005-12-07, 2005-12-20, 2005-12-27, and 2006-01-13 using the Green Bank Telescope (GBT) with the dual-beam dual-polarization receiving system.
We followed the standard spectral calibration procedure using GBTIDL to obtain spectra with a velocity resolution of 0.33 km s$^{-1}$.
The rms noise of the spectrum at each epoch was 2.3, 2.7, 2.1, 1.4, 4.6, and, 2.5 mJy, respectively.
The continuum flux density was 1.451, 1.131, 1.365, 1.272, 1.505, and 1.168 Jy, respectively, involving systematic errors of $\sim 10$\% in flux scaling and off-source subtraction.

\subsection{ALMA}
The Atacama Large Millimeter/submillimeter Array (ALMA) observations with 41 12-m antennas were carried out on 2022-05-12, as the PI science project 2021.1.00341.S targeting a high-redshift galaxy.
The operations software selected NGC 1052 as a phase calibrator within $15^{\circ}$ from the target and J$0423-0120$ as a bandpass calibrator.
On-source integration were 30 s $\times 4$ scans and a 912 s scan, respectively.
Four spectral windows (SPWs) with a 1875-MHz bandwidth were centered at 306.4, 308.2, 318.4 and 320.2 GHz.
The spectral resolution was 7.8125 MHz corresponding to 7.3 km s$^{-1}$.
The image rms was 3.3 mJy/beam/ch with the synthesized beam of $0^{\prime \prime}.68 \times 0^{\prime \prime}.56$ with PA$=79^{\circ}$

\begin{figure}
 \begin{center}
 \includegraphics[width=0.8\linewidth, height=0.8\textheight]{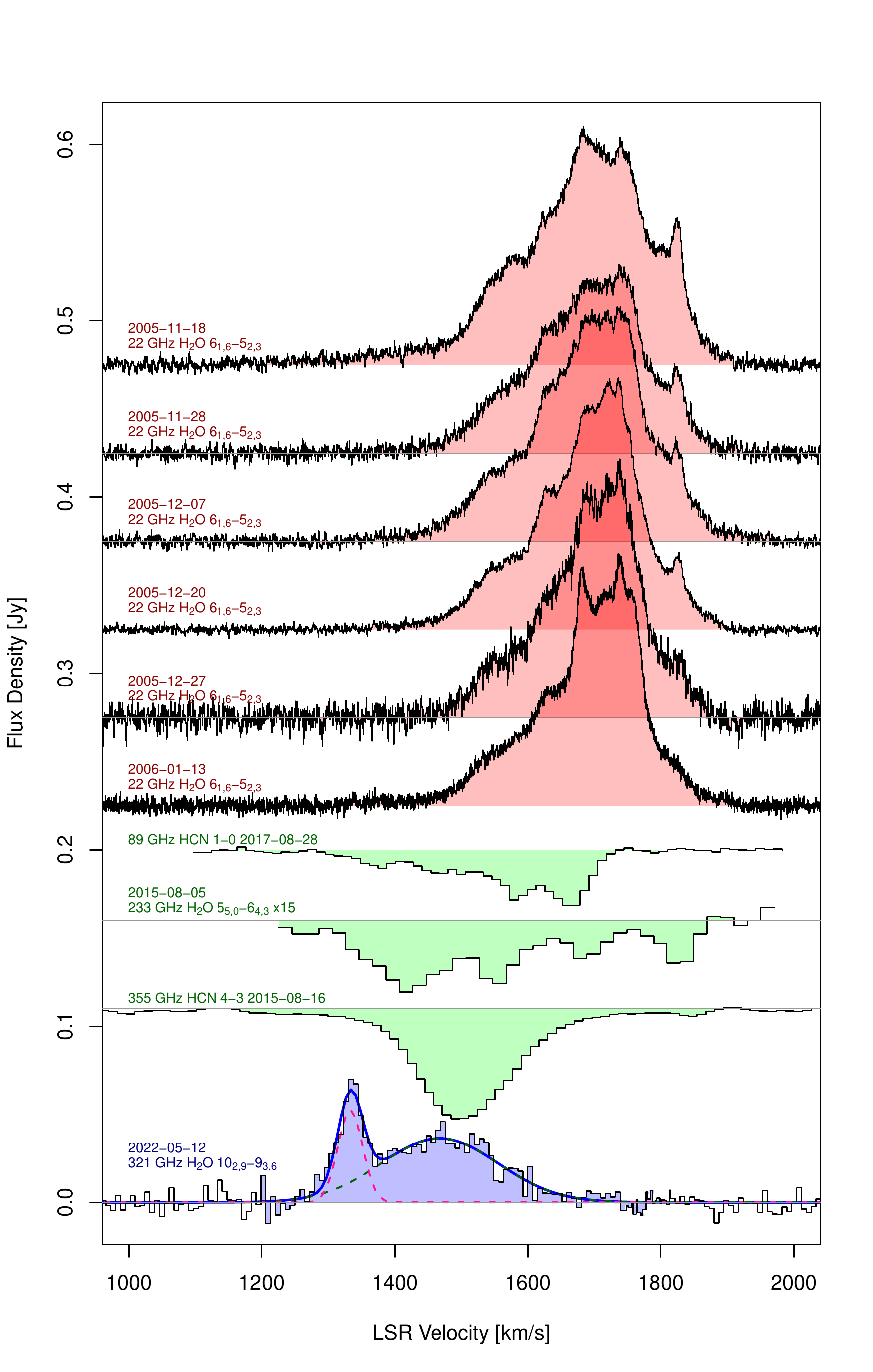}
 \end{center}
 \caption{Continuum-subtracted spectra of NGC 1052. Observation date and line species are labeled on the left side.
The bottom blue spectrum is the 321-GHz H$_2$O emission observed with ALMA. Two adjacent SPWs overlap in the velocity range of 1752 -- 1814 km s$^{-1}$ to show non-uniform channel spacing. The profile is fitted (blue solid line) by two Gaussian components (dashed lines): (a) a broad component with the center velocity of $1467 \pm 5$ km s$^{-1}$ and the width (FWHM) of $208 \pm 12$ km s$^{-1}$ in green, and (b) a narrow component with the center velocity of $1333 \pm 1$ km s$^{-1}$ and the width (FWHM) of $44 \pm 3$ km s$^{-1}$ in red.
Green spectra are the absorption features of 89-GHz HCN $(1-0)$ \citep{2023ApJsubmitted}, 233-GHz H$_2$O $5_{5,0} - 6_{4,3}$, $v_2=1$, and 355-GHz HCN $(4-3)$ \citep{2020ApJ...895...73K} transitions, offset by 0.11, 0.16, and 0.2 Jy, respectively. The 233-GHz feature is vertically magnified by 15 times.
Six red spectra are the 22-GHz masers observed with GBT, offset by $0.525 - 0.05i$ Jy for $i$-th epoch.
The vertical dashed line indicates the systemic velocity of 1492 km s$^{-1}$.}\label{fig:maser_series}
\end{figure}

\begin{figure}
 \begin{center}
  \includegraphics[width=\linewidth]{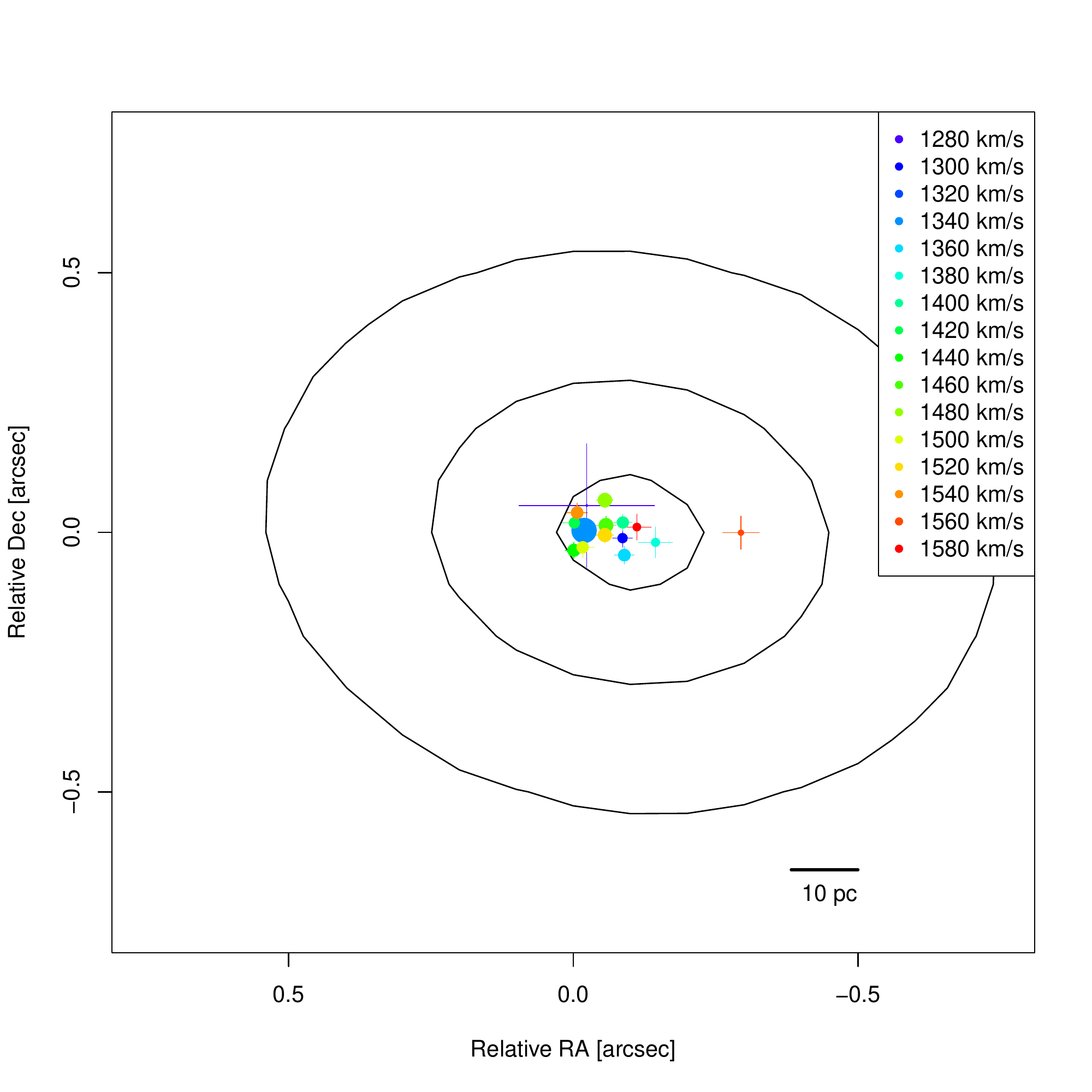}
 \end{center}
 \caption{The contours show continuum image at 313.3 GHz combining all SPWs.
 The contour levels stand for 0.1, 0.5, and 0.9 $\times$ the peak intensity of 0.39 Jy beam$^{-1}$.
 Note that the continuum source is unresolved (see Fig.\ref{fig:uvplot}) and the contours represent the beam size.
Filled circles with crosses indicate continuum-subtracted 321-GHz maser spots with a velocity slice of 20 km s$^{-1}$.
The circle size is proportional to the flux density of the emission line.
Crosses indicate $3\times $ standard errors of the position determined by 2-D Gaussian fit.
The origin of the map is $(\alpha, \delta) = (02^h 41^m 04^s.798, \ -08^{\circ} 15^{\prime} 20^{\prime \prime}.751)$ (J2000).}\label{fig:maserMap}
\end{figure}

\begin{figure}
 \begin{center}
  \includegraphics[width=\linewidth]{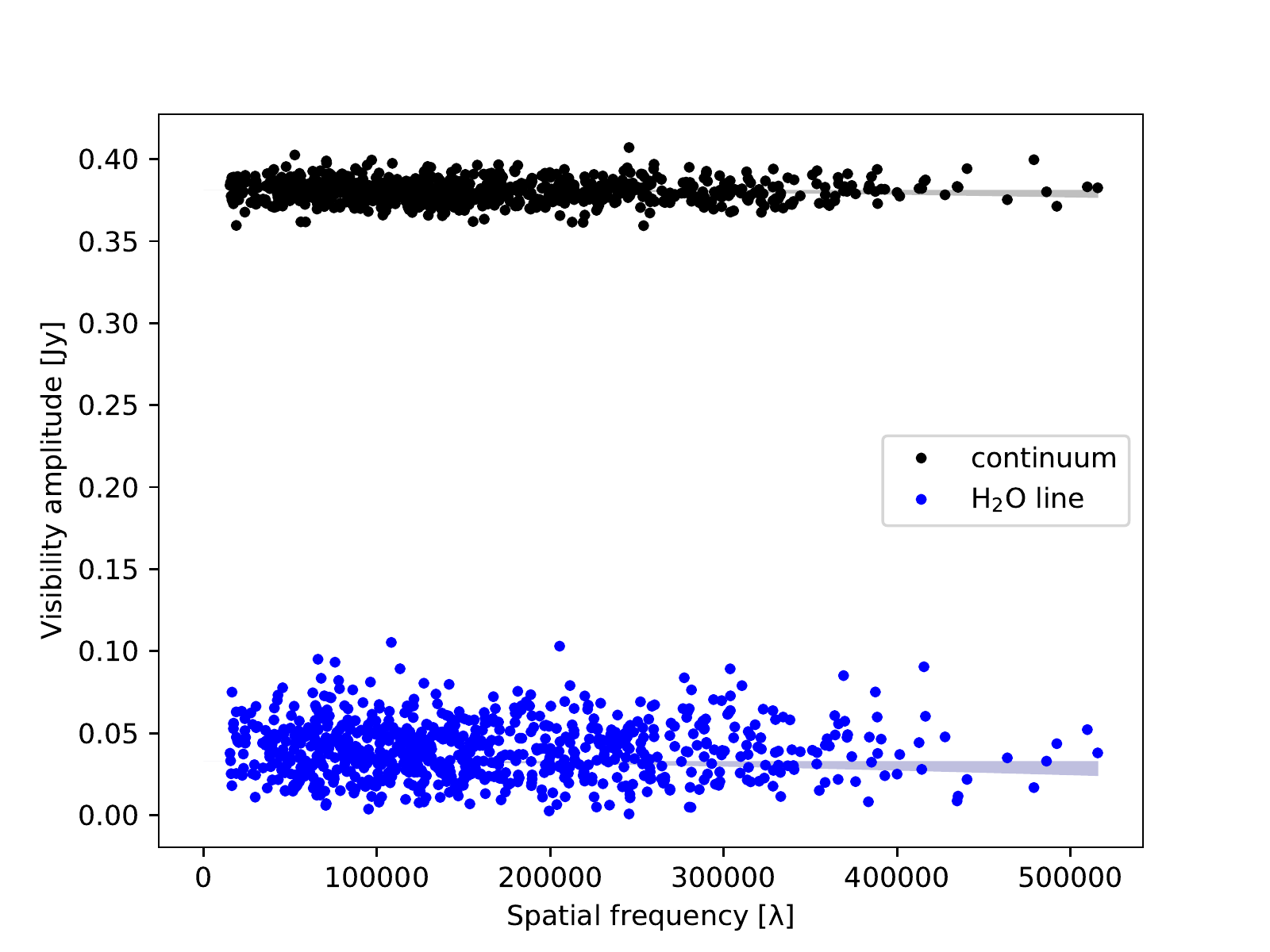}
 \end{center}
 \caption{Visibility amplitudes of 319.3-GHz continuum and line emission as a function of spatial frequency (projected baseline length / wavelength). The 321-GHz H$_2$O line visibility is obtained by integrating the continuum-subtracted spectra in the velocity range of $1300-1550$ km s$^{-1}$. The gray and pale blue filled areas indicate the 99\%-confidence sizes of $\leq 0^{\prime \prime}.024$ and $\leq 0^{\prime \prime}.12$, respectively.}\label{fig:uvplot}
\end{figure}

\section{Results}
Continuum-subtracted spectra are shown in Fig.\ref{fig:maser_series} together with absorption lines of H$_2$O $5_{5,0}-6_{4,3}$, $v_2=1$, HCN $1-0$, and $4-3$ \citep{2020ApJ...895...73K,2023ApJsubmitted}.
The 22-GHz maser showed broad spectra in the velocity range of $1450 - 1850$ km s$^{-1}$ with time-variable narrower components. The centroid of the 22-GHz emission remained at $1685 \pm 4$ km s$^{-1}$, which was redshifted with respect to the systemic velocity by 193 km s$^{-1}$.
The 321-GHz H$_2$O emission profile was biased blueward, which was fitted by two Gaussian components: (a) a broad component with the center velocity of $1467 \pm 5$ km s$^{-1}$ and FWHM of $208 \pm 12$ km s$^{-1}$, and (b) a narrow component with the center velocity of $1333 \pm 1$ km s$^{-1}$ with FWHM of $44 \pm 3$ km s$^{-1}$.
The peak flux density was 0.07 Jy at 1334 km s$^{-1}$.

We produced channel maps of the 321-GHz H$_2$O line with a velocity slice of 20 km s$^{-1}$ in $1280$ km s$^{-1} \leq V_{\rm LSR} \leq 1580$ km s$^{-1}$ and measured the position by 2-D Gaussian fitting in the channel maps.
Fig.\ref{fig:maserMap} shows the position of each channel registered on the continuum image.
These coincided with the continuum position with an upper limit of $0^{\prime \prime}.14$ corresponding to 12 pc, except one outlier offset by $0^{\prime \prime}.29 \pm 0^{\prime \prime}.03$ at 1560 km s$^{-1}$ with a signal-to-noise ratio as low as 4.3.
No significant velocity gradient was found.

The continuum component was unresolved with flux densities of 0.401 Jy and 0.380 Jy at 307.3 GHz and 319.3 GHz, respectively.
The upper limits (99\% confidence) of the FWHM size were $0^{\prime \prime}.024$ and $0^{\prime \prime}.12$ for the continuum and the continuum-subtracted velocity-integrated emission line, respectively, as shown in Fig.\ref{fig:uvplot}.

\section{Discussion}
\subsection{Justification for maser emission} \label{subsec:maseremission}
The 321-GHz H$_2$O emission size of FWHM $\leq 0^{\prime \prime}.12$ and peak flux density of 0.07 Jy yields a brightness temperature of $T_{\rm B} \geq 43$ K.
The ALMA array configuration was too compact ($<500.2$ m) to rule out the possibility of thermal emission.
Nevertheless, ancillary evidence indicates that the emission is non-thermal maser. 

All thermal molecular transitions, including H$_2$O $5_{5,0}-6_{4,3}$, $v_2=1$, appeared in absorption straddling the systemic velocity at the core position of NGC 1052 \citep{2020ApJ...895...73K}.
CO $J=2-1$ and $3-2$ appeared as the sole thermal emission in the circum-nuclear disk (CND) extended in $230 \times 60$ pc off the continuum core, while they appeared in absorption toward the core \citep{2020ApJ...895...73K}.
The 321-GHz emission locates within 12 pc from the radio continuum, much more compact than the CND.
The integrated flux density of the 321-GHz H$_2$O emission was 10.5 Jy km s$^{-1}$ exceeding the 1.55 Jy km s$^{-1}$ and 7.96 Jy km s$^{-1}$ of CO emissions.
The CND cannot harbor the 321-GHz H$_2$O emission.

Since the brightness temperature of the continuum emission exceeds $10^8$ K \citep{2004A&A...426..481K}, molecular gas covering the continuum source must be seen in absorption except non-thermal emission.
Thus, the 321-GHz H$_2$O emission at the same position of the core is probably non-thermal maser.
Longer-baseline ($> 3500$ m) observations would provide more concrete constraints on the brightness temperature $> 2000$ K.

\subsection{Spectral profiles and location of masers} \label{subsec:location}
Observed maser intensity is a product of the background brightness and amplification gain through population-inverted H$_2$O molecules.
A line width becomes progressively narrower than the thermal width as it gains more amplification \citep{2003adu..book.....D}.
The wide velocity width in NGC 1052 indicates small amplification gain and requires bright background continuum emission.
Thus, the position of maser spots is related to the continuum structure.
That characteristic is consistent with the smooth systemic-velocity components and is different from narrow spikes in archetypal megamasers.

\citet{2008ApJ...680..191S} revealed that the position of 22-GHz masers in NGC 1052 coincides with the intensity peaks of the double-sided jets where the FFA opacity is not too high.
The continuum core is hidden in the optically-thick gap at 22 GHz to generate 22-GHz maser avoidance.
The GBT monitoring of 22-GHz maser emission always shows redshifted profiles with respect to the systemic velocity, implying an inward stream of the excited H$_2$O molecule inside the near side of the torus.

The continuum structure at higher frequency is more dominated by a compact core with a smaller contribution from steep-spectrum jets, and the gap due to FFA disappears in millimeter (mm) wavelengths \citep{2016ApJ...830L...3S,2016A&A...593A..47B,2019A&A...623A..27B,2019ApJ...872L..21S}.
Thus, the 321-GHz maser is considered to pinpoint the compact core and is amplified by population-inverted H$_2$O molecules with a temperature $>1000$ K.
\citet{2020ApJ...895...73K} found sub-mm molecular absorption features (see HCN $4-3$ in  Fig.\ref{fig:maser_series}) that implied presence of a massive molecular torus covering the continuum source.
The velocity ranges of the sub-mm absorption lines straddle the systemic velocity and resemble that of the broad component of the 321-GHz maser.
This supports that molecular gas in the torus causes absorption and stimulated emission along the line of sight to the compact core.

\citet{2023ApJsubmitted} found mm (86 -- 130 GHz) molecular absorption features and identified different line profiles from sub-mm ones (see HCN $1-0$ and $4-3$ in Fig.\ref{fig:maser_series}).
The difference is ascribed to the continuum structure dominated by the jet at mm and the core at sub-mm as mentioned in the former paragraph.
The temperature of SO absorption lines were $26 \pm 4$ K and $344 \pm 43$ K for mm and sub-mm, respectively. The difference in temperatures also indicate that mm and sub-mm absorption lines represent different absorbers.

The torus is expected to consist of multiple layers with temperatures of $\geq 10^4$ K, $\geq 1000$ K, $\geq 400$ K, and $344 \pm 43$ K that accounts for the FFA, 321-GHz maser, 22-GHz maser, and sub-mm SO absorption region, respectively.
\citet{2023ApJsubmitted} interpret that SO molecules evaporate from dust grains through jet-torus interaction and the high temperature is maintained by shock heating.
The shock heating also explains the excitation of the 321-GHz maser. 

Since the mm SO absorber is too cold to excite the maser, downstream of the jets is unlikely to harbor the 321-GHz maser.

\subsection{Comparison with the Circinus galaxy and NGC 4945}
In this subsection we compare the properties of the H$_2$O maser in NGC 1052 with those in NGC 4945 and the Circinus galaxy referring to \citet{2021ApJ...923..251H}.

The isotropic luminosity is estimated to be 176 $L_{\Sol}$ and 1090 $L_{\Sol}$ for 22-GHz and 321-GHz masers, respectively, using the formula $\displaystyle L = 1.04 \times 10^{-3} \nu_{\rm rest}, D^2_{L} \int S(v) \ dv$.
These values are 5.5 times and 17 times larger than those of the Circinus galaxy.
The ratio of the 321-GHz maser peak flux to continuum flux in NGC 1052 is 0.19, which is significantly smaller than that of 3.2 and 37 in the Circinus galaxy on 2012-06-03 and 2017-05-06, respectively.
The ratio in NGC 4945 was 0.29 on 2012-06-03 and there was a non detection on 2017-05-06.
Since the ratio is an indicator of maser amplification gain, the high isotropic luminosity in NGC 1052 is ascribed to the brighter continuum.
The broader maser feature is consistent with less narrowing with the lower amplification gain. It is difficult to establish the degree of blending of multiple narrower features making up the spectrum in these observations however. 

While the Circinus galaxy and NGC 4945 harbor high-velocity components $\sim 1000$ km s$^{-1}$ with respect to the systemic velocity, such components are not identified in NGC 1052.
\citet{1995PNAS...9211427M} interpreted the high-velocity components of 22-GHz masers as self-amplification of long velocity-coherent gain path through the rotating disk without a background continuum source.
Presence of the systemic-velocity component and absence of high-velocity components in NGC 1052 also support lower gain amplification of  brighter continuum.

\section{Summary}
ALMA observations of NGC 1052 revealed the presence of 321 GHz H$_2$O emission with the isotropic luminosity of 1090 $L_{\Sol}$ which is in all probability the first sub-mm maser in a radio galaxy.
The sub-mm maser profile straddles the systemic velocity of the galaxy, similar to sub-mm molecular absorption lines, unlike the redshifted profile of the 22-GHz maser.
Although the sub-mm maser distribution is unresolved, the inferred location is between the two clusters of 22-GHz masers pinpointing the compact continuum core.

Followup ALMA observations of the 321 GHz emission, in order to establish variability and polarization in that line, as well as searching for emission in other H$_2$O maser lines such as at 183 and 325 GHz, could also provide valuable information on the nucleus of this target.
Sub-mm VLBI observations are desired to resolve the structure of the maser and unveil the dynamics and the excitation mechanism in the sub-pc region of AGN.

\bigskip
This letter makes use of the following GBT and ALMA data: GBT 05C-034 and ADS/JAO.ALMA\#2021.1.00341.S. The Green Bank Observatory is a facility of the National Science Foundation operated under cooperative agreement by Associated Universities, Inc. ALMA is a partnership of ESO (representing its member states), NSF (USA) and NINS (Japan), together with NRC (Canada), MOST and ASIAA (Taiwan), and KASI (Republic of Korea), in cooperation with the Republic of Chile. The Joint ALMA Observatory is operated by ESO, AUI/NRAO and NAOJ.
This work is supported by JSPS KAKENHI 18K03712 and 21H01137.

\end{document}